\documentclass[aps,prl,superscriptaddress,twocolumn,notitlepage]{revtex4-1}
\usepackage[dvips]{graphicx}
\usepackage{dcolumn}
\usepackage{bm}
\usepackage{amsmath,amsthm,amssymb}
\usepackage{epstopdf}
\usepackage{color}
\usepackage{bbold}
\usepackage{mathtools}
\usepackage[sort&compress]{natbib}
\usepackage{hyperref}
\hypersetup{colorlinks=true,linkcolor=blue,citecolor=blue,urlcolor=black}

\definecolor{azure}{rgb}{0.0, 0.5, 1.0}
\definecolor{amber}{rgb}{1.0, 0.49, 0.0}
\definecolor{red}{rgb}{1.0, 0.1, 0.0}
\definecolor{forestgr}{rgb}{0.13, 0.55, 0.13}

\newcommand{\bra}[1]{\left<#1\right|}
\newcommand{\ket}[1]{\left|#1\right>}
\newcommand{\mbf}{\mathbf}

\newcommand{\D}{\hat{\mathcal{D}}}

\renewcommand{\Re}{\mathrm{Re}\,}
\renewcommand{\Im}{\mathrm{Im}\,}

\begin{document}

\title{Dark states of multilevel fermionic atoms in doubly-filled optical lattices}

\author{A. Pi\~neiro Orioli}
\affiliation{JILA, NIST, Department of Physics, University of Colorado, Boulder, CO 80309, USA}
\affiliation{Center for Theory of Quantum Matter, University of Colorado, Boulder, CO 80309, USA}

\author{A. M. Rey}
\affiliation{JILA, NIST, Department of Physics, University of Colorado, Boulder, CO 80309, USA}
\affiliation{Center for Theory of Quantum Matter, University of Colorado, Boulder, CO 80309, USA}

\pacs{}
\date{\today}

\begin{abstract}
We propose to use fermionic atoms with degenerate ground and excited internal levels ($F_g\rightarrow F_e$), loaded into the motional ground state of an optical lattice with two atoms per lattice site, to realize dark states with no radiative decay. The physical mechanism behind the dark states is  an interplay of Pauli blocking and multilevel dipolar interactions. The dark states are independent of lattice geometry, can support an extensive number of excitations and can be coherently prepared using a Raman scheme taking advantage of   the quantum Zeno effect. These attributes make them appealing for  atomic clocks,  quantum memories, and   quantum information on decoherence free subspaces.
\end{abstract}

\maketitle

%%%%%%%%%%%%%					%%%%%%%%%%%%%
%%%%%%%%%%%%%		SECTION		%%%%%%%%%%%%%
%%%%%%%%%%%%%					%%%%%%%%%%%%%

{\it Introduction.---}Subradiance is a fascinating quantum phenomenon in which coupled quantum emitters, e.g.~excited atoms, radiate light at a slower rate than independent emitters.
This modified decay rate originates from quantum interference between single-particle and collective decay processes, due to atom-light interactions.
When the decay rate completely vanishes, the corresponding states are called \emph{dark} states.

Ever since Dicke's seminal paper~\cite{DickePR93}, subradiance has been widely studied in two-level systems~\cite{ScullyPRL115,RuostekoskiPRL117,RitschSciRep2015,AsenjoPRA95,AsenjoPRX2017,AsenjoAlbrechtNJP2019,ChangPRA99,RitschCardoner2019,ZollerPRL122,RobicheauxPRA94,JenPRA96,TudelaPRL115,LukinPRL119,ShahmoonPRL118,LesanovskyBuonaiuto2019,LesanovskyPRA97},
and in systems with multiple excited levels decaying to a unique ground state~\cite{EversPRA75,Ritsch1905.01483,LesanovksyNeedham2019,LesanovskyPRL110}.
Most works concentrate on subradiant states where a single excitation is shared among all atoms.
Recently, single-excitation subradiant states in arrays of two-level atoms have been shown to have interesting applications for quantum memories~\cite{AsenjoPRX2017}, atomic clocks~\cite{ChangPRA99}, mirrors~\cite{ShahmoonPRL118}, excitation transport~\cite{RitschCardoner2019,LesanovksyNeedham2019}, or to create topological states~\cite{LukinPRL119,Syzranov2016} or entangled photons~\cite{TudelaPRL115}.
However, despite many theoretical proposals, only few experiments have managed to observe subradiance so far~\cite{KaiserPRL116,TemnovPRL95,SolanoNatComm2017,ZhouNature2011,DeVoePRL76,EschnerNature2001,HettichScience2002,ZelevinskyNatPhys2015,JulienneTakasuPRL108,WallraffScience2013}.
This is challenging because subradiant states are generally hard to prepare, and often require the atoms to be very close to each other compared to the wavelength of the transition.

\begin{figure}[t!]
\centering
	\includegraphics[width=\columnwidth]{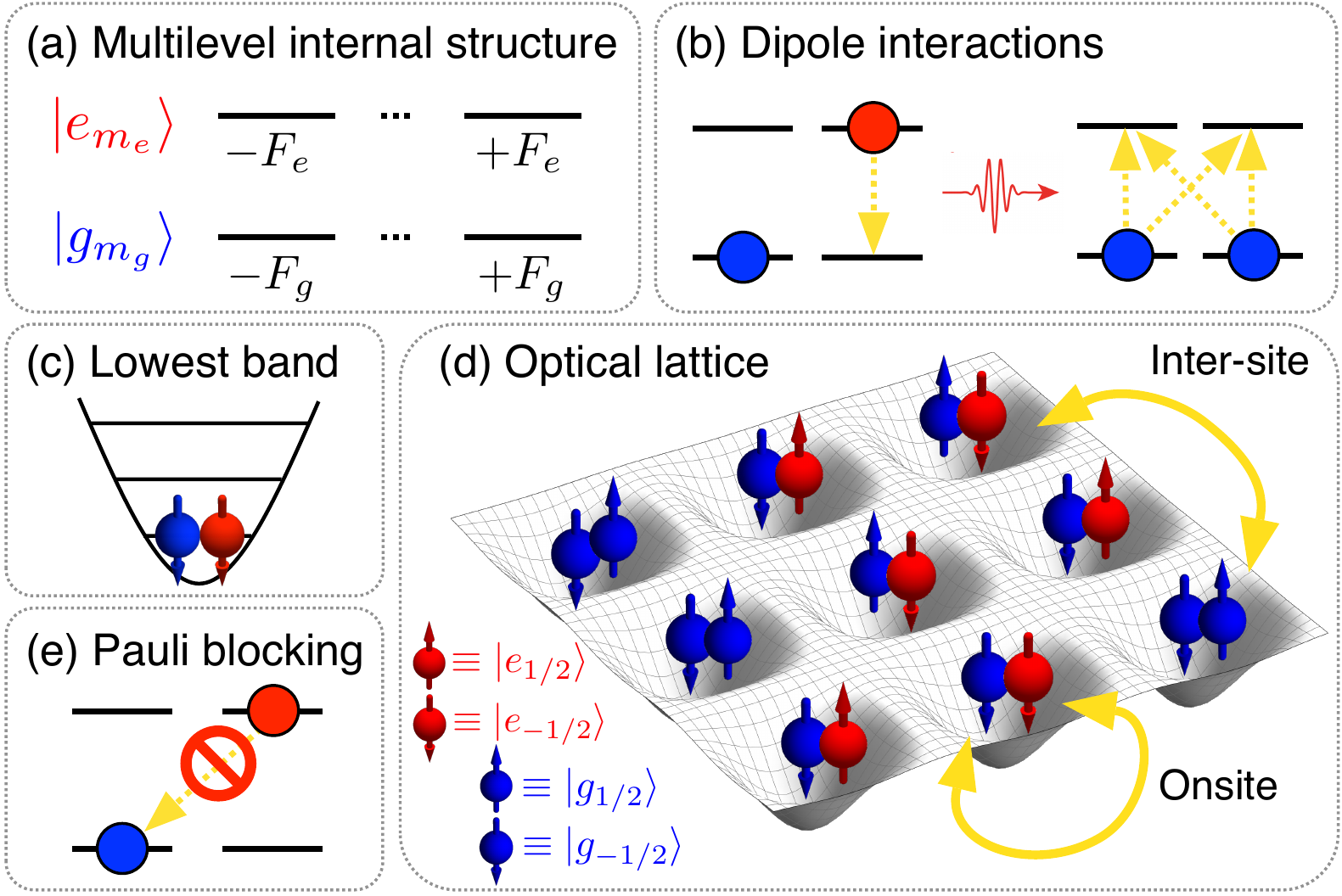}
	\caption{Summary of proposed platform. (a) Fermionic atoms with an $F_g\rightarrow F_e$ transition from ground (blue) to excited levels (red). (b) Examples of dipole exchange interaction between different levels mediated by a photon (red wave). (c),(d) Optical lattice with two atoms per site occupying the motional ground state only.  (e) Example of a decay channel blocked by Pauli exclusion.
	}
	\label{fig:platform}
\end{figure}

In this work, we propose to circumvent some of these problems using fermionic atoms with multiple internal levels.
Such atoms possess both degenerate ground and excited levels.
Due to the complexity of the problem, only few recent works have studied subradiance in multilevel systems with degenerate ground states~\cite{RitschPRL118,Asenjo1906.02204}.
In particular, the multiple decay channels available to each excited state make it hard to generate subradiant states.
Here, however, we take advantage of the blockade imposed by fermion statistics by considering arrays with two fermionic atoms per optical lattice site [Fig.~\ref{fig:platform}(e)].

For this system we find a large set of dark states with remarkable features.
They are independent of lattice geometry, in particular they do not require subwavelength arrays, they can support an extensive number of excitations, and they can be coherently prepared using e.g.~a Raman scheme.
These dark states arise from a combination of having two atoms per site in the motional ground state, Pauli blocking, and multilevel dipolar interactions.
We focus here on implementations with alkaline-earth(-like) atoms such as $^{171}$Yb and $^{87}$Sr, but emphasize that the main results are not restricted to these species.
Our findings open the door to potentially using this decoherence-free subspace of dark states for, e.g., atomic clocks or quantum memories.

%%%%%%%%%%%%%					%%%%%%%%%%%%%
%%%%%%%%%%%%%		SECTION		%%%%%%%%%%%%%
%%%%%%%%%%%%%					%%%%%%%%%%%%%

{\it System.---}We consider an optical lattice loaded with $n=2$ fermionic atoms per lattice site, see Fig.~\ref{fig:platform}. We work in the limit of a deep trap such that tunneling is suppressed, and the atoms occupy the motional ground-state only [Fig.~\ref{fig:platform}(c)].
The latter is valid as long as the onsite trapping frequency is much larger than the photon recoil energy (Lamb-Dicke regime), and typical atom-atom interactions.

We consider for each atom a radiative transition with half-integer total angular momentum $F_g\rightarrow F_e$.
The internal level structure thus consists of a manifold of ($2F_g+1$)-degenerate ground states, $\ket{g_{m_g}}\equiv\ket{g,F_g,m_{g}}$ with $m_g\in[-F_g,F_g]$, and a manifold of ($2F_e+1$)-degenerate excited states, $\ket{e_{m_e}}\equiv\ket{e,F_e,m_{e}}$ with $m_e\in[-F_e,F_e]$ [Fig.~\ref{fig:platform}(a)].
These two sets of states are separated by an energy $\omega_0=ck_0$.

The atoms interact with each other via dipole interactions (Fig.~\ref{fig:platform}).
After a standard Born-Markov approximation~\cite{GrossHarochePRep1982,JamesPRA47,LehmbergPRA2}, the dynamics of the atomic density matrix $\hat{\rho}$ can be described by a master equation, $\dot{\hat{\rho}}=-i[\hat{H},\hat{\rho}] + \mathcal{L}(\hat{\rho})$ ($\hbar=1$), with
\begin{align}
	\hat{H} =&\, - \sum_{i,j} \sum_{q,q'} \mathcal{R}_{q,q'}^{i,j}\, \D^+_{i,q}\, \D^-_{j,q'} ,
\label{eq:HLdipoles}\\
	\mathcal{L}(\hat{\rho}) =&\, - \sum_{i,j} \sum_{q,q'} \mathcal{I}_{q,q'}^{i,j} \left( \left\{ \D^+_{i,q}\, \D^-_{j,q'}, \hat{\rho} \right\} - 2 \D^-_{j,q'}\, \hat{\rho}\, \D^+_{i,q} \right).\nonumber
\end{align} 
This describes all possible coherent and incoherent exchanges of photons between two atoms.
The crucial element here is the spherical dipole operator $\D^\pm_{i,q}$, which acts as a multilevel raising and lowering operator defined as $\hat{\mathcal{D}}^+_{i,q}\equiv(\hat{\mathcal{D}}^-_{i,q})^\dagger$ and
\begin{equation}
	\D^-_{i,q} = \sum_{m} C^q_m\, \hat{\sigma}^{(i)-}_{m,m+q}.
\label{eq:D-def}
\end{equation}
Here, $\hat \sigma^{(i)-}_{mn}\equiv \hat f^\dagger_{i,g_{m}}\hat f_{i,e_{n}}$, and $\hat f^{(\dagger)}_{i,a_m}$ annihilates (creates) a fermion at site $i$ with internal level $\ket{a_m}$ ($a=g,e$), and $\{ \hat f_{i,a_m} , \hat f^\dagger_{j,b_n} \} = \delta_{ij}\,\delta_{ab}\,\delta_{mn}$.
Thus, the operator $\D^-_{i,q}$ corresponds to a sum over all possible decay processes from $\ket{e_{m+q}}$ to $\ket{g_m}$, weighted by the Clebsch-Gordan coefficient of the transition, $C^q_m\equiv \langle F_g,m;1,q | F_e,m+q \rangle$. The emitted photon can have polarization $q=0,\pm1$ with $\mbf e_0\equiv\mbf e_z$ and $\mbf e_\pm\equiv\mp(\mbf e_x\pm i\mbf e_y)$, where $\mbf e_z$ defines the quantization axis.
The strength of the interaction, $\mathcal{R}_{q,q'}^{i,j}\equiv \left( \mbf e_q^{*T}\, \Re G^{ij}\, \mbf e_{q'} \right)$ and $ \mathcal{I}_{q,q'}^{i,j}\equiv  \left( \mbf e_q^{*T}\, \Im G^{ij}\, \mbf e_{q'} \right)$, depends on the polarizations $q$ and $q'$ of the involved transitions and on the relative distance between the atoms.

For atoms at different sites ($i\neq j$), the dipolar interaction coefficients can be written as $G^{ij} = G(\mbf r_i-\mbf r_j)$, where $G$ is proportional to the electromagnetic dyadic Green's tensor in vacuum~\cite{novotny2006principles}
$G(\mbf r) = \frac{3\Gamma}{4} \left\{ \big[ \mathbb{1} - \hat{\mbf r}\otimes \hat{\mbf r} \big] \frac{e^{ik_0r}}{k_0r}+ \big[ \mathbb{1} - 3\, \hat{\mbf r}\otimes \hat{\mbf r} \big] \left( \frac{ie^{ik_0r}}{(k_0r)^2} - \frac{e^{ik_0r}}{(k_0r)^3} \right) \right\}.$ Here, $\hat{\mbf{r}}\equiv \mbf r/|\mbf r|$, and the spontaneous decay rate is defined as $\Gamma = \omega_0^3|d^{\text{rad}}_{ge}|^2/[3\pi\epsilon_0\hbar c^3(2F_e+1)]$, where $d^{\text{rad}}_{ge}$ is the radial dipole matrix element~\cite{Supplemental}.
The onsite ($i=j$) interaction coefficients involve an integral of the dyadic Green's tensor over the spatial part of the wave-function~\cite{Supplemental}.
In the limit of a deep, radially symmetric trap potential for the two atoms on the same lattice site, they can be approximated by $\Re G^{ii}=0$ and $\Im G^{ii}=\frac{\Gamma}{2} \mathbb{1}$.
Nevertheless, we emphasize that our results are independent of the trap details and the specific form of $G^{ii}$.

%%%%%%%%%%%%%					%%%%%%%%%%%%%
%%%%%%%%%%%%%		SECTION		%%%%%%%%%%%%%
%%%%%%%%%%%%%					%%%%%%%%%%%%%

{\it Multilevel dark states.---}Mathematically, a dark state $\ket{D}$ is defined as an eigenstate of $\hat{H}$ with $\mathcal{L}(\ket{D}\!\bra{D})=0$.
From Eq.~(\ref{eq:HLdipoles}), a sufficient condition to fulfill this is given by
\begin{equation}
	\D^-_{i,q} \ket{D} = 0\quad \forall i,q.
\label{eq:dark_condition}
\end{equation}
Physically, this means that all possible photon emission processes, or \emph{decay processes}, of the state $\ket{D}$ need to interfere destructively. Specifically, condition~(\ref{eq:dark_condition}) requires that each possible polarization $(q=0,\pm1)$ cancels out independently from the others.
On top of this, interference of different decay processes can only happen if the final state $\ket{f}$ is the same.
Since excited states can decay in our case to different ground states, this implies that all possible decay processes of $\ket{D}$ with polarization $q$ and final state $\ket{f}$ have to cancel out independently from the other polarizations and final states.
Hence, each possible pair $(q,\ket{f})$, or \emph{decay channel}, gives rise to a separate condition to be fulfilled.
Notice that all channels are nevertheless intertwined in a complex fashion since each state has, in general, multiple decay channels.
The situation is, thus, far more complex than for two-level systems, which have only one relevant polarization and ground-state.

It is this multilevel complexity, however, that allows to find solutions to Eq.~(\ref{eq:dark_condition}) when combined with the other two key ingredients of our proposal: two atoms per site and fermion statistics.
These two elements essentially allow to block certain unique decay channels which could otherwise not be cancelled out by interference.
To see this, we first consider the case of a single lattice site, which will later allow us to construct dark states for the multisite system.

\begin{figure}[t!]
\centering
	\includegraphics[width=\columnwidth]{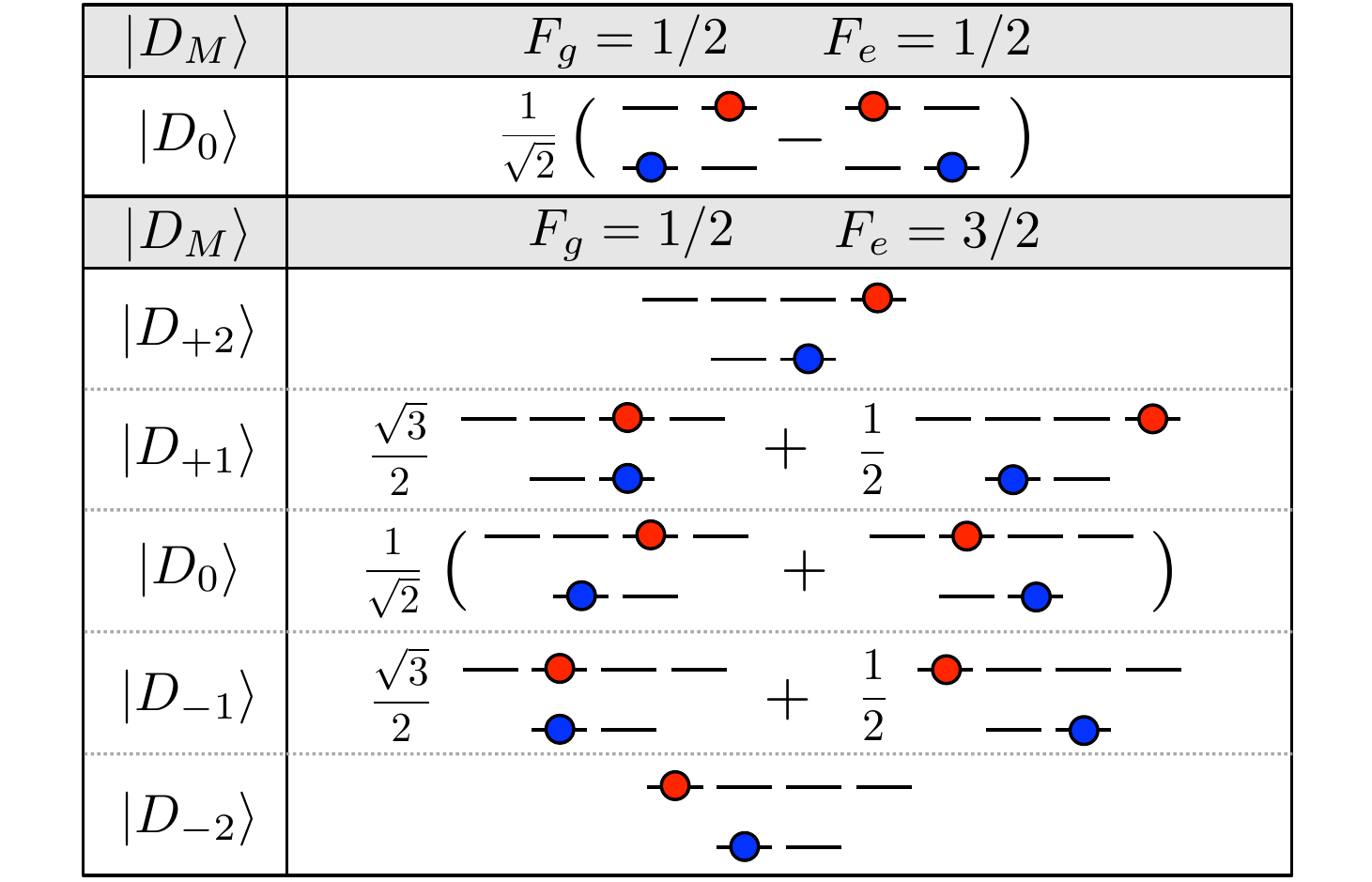}
	\caption{Dark states for two fermions on a single site for different internal level structures.}
	\label{fig:dark_states}
\end{figure}

As a specific example, we consider two atoms on a single lattice site with $F_g=F_e=1/2$.
In this case, there exists exactly one dark state given by (c.f.~Fig.~\ref{fig:dark_states}) $ \ket{D_0}_{\{\frac{1}{2},\frac{1}{2}\}} \equiv\frac{1}{\sqrt{2}} \left( \ket{g_{-1/2}\,e_{1/2}} - \ket{g_{1/2}\,e_{-1/2}} \right)$,  where we defined Fock states as $\ket{a_m b_n} \equiv \hat{f}^\dagger_{a_m} \hat{f}^\dagger_{b_n} \ket{\text{vacuum}}$ with $a,b\in\{g,e\}$.
The darkness of this state relies heavily on the Pauli exclusion principle. Because of it, each of the Fock states involved in the superposition can only decay to $\ket{g_{-1/2}g_{1/2}}$, via $q=0$ polarization. These two contributions interfere destructively.
However, if the atoms could occupy the same state, then, e.g., $\ket{g_{1/2}\,e_{-1/2}}$ could decay to $\ket{g_{1/2}\,g_{1/2}}$, which could not be cancelled out by interference.

Notice that the dark state  $\ket{D_0}_{\{\frac{1}{2},\frac{1}{2}\}}$ is symmetric in $e$ and $g$, as opposed to the usual two-level Dicke dark state $(\ket{eg}-\ket{ge})/\sqrt{2}$. The reason for this lies in the properties of the Clebsch-Gordan coefficients, which determine the amplitudes of the states involved in the dark states, here specifically $C^{q=0}_{1/2}=-C^{q=0}_{-1/2}$.

Another illustrative example of this is given by the level structure $F_g=1/2 \rightarrow F_e=3/2$. In this case, there exist three dark states given by (c.f.~Fig.~\ref{fig:dark_states}) $\ket{D_0}_{\{\frac{1}{2},\frac{3}{2}\}} \equiv  \frac{1}{\sqrt{2}} \left( \ket{g_{-1/2}\,e_{1/2}} + \ket{g_{1/2}\,e_{-1/2}} \right)$ and 
$\ket{D_{\pm1}}_{\{\frac{1}{2},\frac{3}{2}\}} \equiv\frac{\sqrt{3}}{2} \ket{g_{\pm1/2}\,e_{\pm1/2}} + \frac{1}{2} \ket{g_{\mp1/2}\,e_{\pm3/2}} $,
where the subscript stands for the total $M=m_g+m_e$ of the state, $\ket{D_{M}}$. For this internal level structure one finds that $C^{q=0}_{1/2}=C^{q=0}_{-1/2}$ and, hence, the state $\ket{D_0}_{\{\frac{1}{2},\frac{3}{2}\}}$ is instead antisymmetric in $e$ and $g$. In comparison, the states $\ket{D_{\pm1}}_{\{\frac{1}{2},\frac{3}{2}\}}$ are asymmetric with prefactors related to Clebsch-Gordan coefficients, as explained below.
Apart from these, there are also two other dark states, $\ket{g_{1/2}\,e_{3/2}}$ and $\ket{g_{-1/2}\,e_{-3/2}}$, which are trivially dark due to Pauli blocking. Such states are in general hard to prepare and will not be further discussed.

This can be generalized to other internal level structures.
For a generic level structure with $F_g=F_e\equiv F$, we find a single dark state given by
\begin{equation}
	\ket{D_0}_{\{F,F\}} \equiv \frac{1}{\sqrt{2F+1}} \sum_{m=-F}^{F} (-1)^{F-m}\, \ket{g_{m}\, e_{-m}}.
\label{eq:dark_n2square}
\end{equation}
For $F=1/2$ this coincides with $\ket{D_0}_{\{\frac{1}{2},\frac{1}{2}\}}$.
More generally, it turns out that this dark state corresponds to an eigenstate $\ket{F_T=0,M=0}$ of the total angular momentum of the two particles, $\mbf{F}_T=\mbf{F}_1+\mbf{F}_2$, where $\mbf{F}_{1/2}$ is the total angular momentum of each atom. This allows for an alternative explanation of the dark state in terms of the symmetries of the orbital ($g,e$) and angular momentum projection ($m_{g/e}$) parts of the wave function (the motional part is symmetric by assumption).
A state with $F_T=0$ can in principle only decay to a state with $F_T=1$.
However, for both atoms in the ground state the state with $F_T=1$ is symmetric~\cite{devanathan2006angular}, and so is the orbital part ($\ket{gg}$).
Such a final state is not allowed by fermion statistics and, therefore, the state of Eq.~(\ref{eq:dark_n2square}) is dark.

For $F_e=F_g+1\equiv F+1$, there exists one dark state, with maximal total angular momentum $F_T=2F+1$, for each possible value of $M\in\{-2F-1,\ldots,2F+1\}$,
\begin{equation}
	\ket{D_M}_{\{F,F+1\}} \equiv \sum_{m=-F+\max(0,M-1)}^{F+\min(0,M+1)} \alpha^{(F,M)}_{m}\, \ket{g_{m}e_{M-m}},
\label{eq:dark_n2V}
\end{equation}
where $\alpha^{(F,M)}_{m}\equiv\langle F,m;F+1,M-m | 2F+1,M \rangle$. These dark states can be understood by noting that the above state with $F_T=2F+1$ can only decay to a state with $F_T=2F$, which is the maximal angular momentum possible for both atoms in the ground state. The $F_T=2F$ state is, however, symmetric~\cite{devanathan2006angular}, and so is the orbital part. Hence, this final state is again not allowed by exchange symmetry, and the state of Eq.~(\ref{eq:dark_n2V}) is dark.

We note that for $F_e=F_g-1$ no dark states exist because the number of decay channels is too large, and that for $|F_e-F_g|\geq2$ the transitions are not dipole allowed.
For fillings $n\geq3$ the number of dark states rapidly increases. Such states are, however, sensitive to three-body losses and will be studied elsewhere~\cite{OrioliPRAinprep}.

We now return to the multisite problem.
Usually, having more than one lattice site introduces new decay channels that makes it harder to form truly dark states.
Remarkably though, for our system a large set of dark states can be built out of simple product states of single-site dark and ground states. More specifically, let $\ket{D_{\alpha_i}}_i$, $i\in\mathcal{S}_D$, be arbitrary $i$-site dark states fulfilling $\D^-_{i,q} \ket{D_{\alpha_i}}_i = 0$ $\forall q$, e.g.~the states of Eqs.~(\ref{eq:dark_n2square}) or (\ref{eq:dark_n2V}).
And let $\ket{G_{\beta_j}}_j$, $j\in\mathcal{S}_G$, be arbitrary $j$-site states with all atoms in the ground-state manifold, which are also trivially dark, $\D^-_{j,q} \ket{G_{\beta_j}}_j = 0$ $\forall q$.
Then, any arbitrary product state given by
\begin{equation}
	\ket{D_{\{\alpha_i\},\{\beta_j\}}} \equiv \bigotimes_{i\in\mathcal{S}_D} \ket{D_{\alpha_i}}_i \bigotimes_{j\in\mathcal{S}_G} \ket{G_{\beta_j}}_j
\label{eq:dark_multisite}
\end{equation}
fulfills Eq.~(\ref{eq:dark_condition}) and is hence a dark state of the multisite system.
Notice that these states are robust against imperfect filling, as long as each site fulfills Eq.~(\ref{eq:dark_condition}), separately.

Dark states of the form (\ref{eq:dark_multisite}) have remarkable properties.
First, they are independent of the geometry of the lattice. In particular they do not require short subwavelength distances between sites, which is usually an important ingredient in proposals for two-level subradiant states.
Second, they can support a large number of excitations, up to one excitation per lattice site.
This contrasts with the two-level case, where typical subradiant states involve only one or few excitations shared among all the atoms, see e.g.~Refs.~\cite{AsenjoAlbrechtNJP2019,ZollerPRL122,RuostekoskiPRL117,RobicheauxPRA94,RitschSciRep2015}.
Moreover, all dark states of the form (\ref{eq:dark_multisite}) with equal number of excitations are degenerate with zero energy shift ($\hat{H}\ket{D}=0$), which opens the door to the creation of stable entangled states.

%%%%%%%%%%%%%					%%%%%%%%%%%%%
%%%%%%%%%%%%%		SECTION		%%%%%%%%%%%%%
%%%%%%%%%%%%%					%%%%%%%%%%%%%

{\it Preparation.---}To excite any multisite dark state of Eq.~(\ref{eq:dark_multisite}) it is sufficient to consider the case of a single lattice site (we drop the index $i$ in the following).
Dark states are typically hard to prepare.
In our case, a laser addressing the $e$-$g$ transition can not coherently excite atoms into the dark state.
This is because the Hamiltonian is given by $\hat{H}_L=-\sum_q\big( \Omega_q\,\D^+_q+\text{h.c.} \big)$ with Rabi frequency $\Omega_{q} \equiv \Omega \left( \mbf e^*_{q} \cdot \boldsymbol\epsilon_L \right)$ and laser polarization $\boldsymbol\epsilon_L$.
This implies $\bra{D}\hat{H}_L\ket{G}=0$ for any ground state $\ket{G}$ and dark state $\ket{D}$ fulfilling Eq.~(\ref{eq:dark_condition}).

\begin{figure}[t!]
\centering
	\includegraphics[width=\columnwidth]{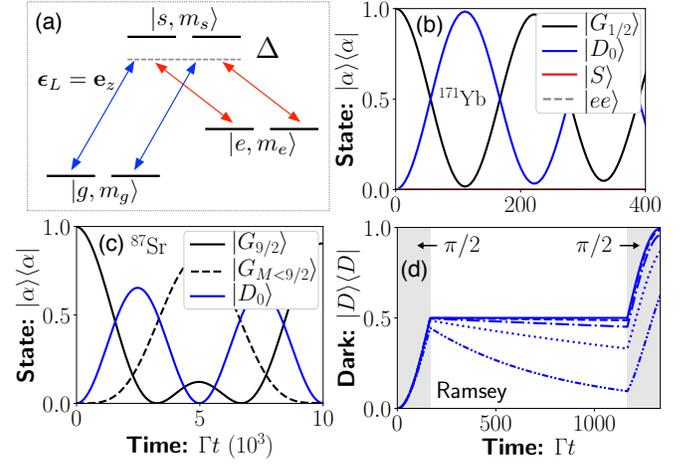}
	\caption{(a) Raman excitation scheme for $F_g=F_e=F_s=1/2$ and $\mbf e_z$ polarized light. (b) Excitation of $\ket{D_0}_{\{1/2,1/2\}}$ for $^{171}$Yb level structure with $\Omega^\text{eff}=0.03\Gamma$. (c) Excitation of $\ket{D_0}_{\{9/2,9/2\}}$ for $^{87}$Sr level structure with $\Omega^\text{eff}=0.001\Gamma$. (d) Resonant Ramsey sequence for $\Delta_z/\Gamma=\{0,4,8,16,32\}10^{-3}$ from top to bottom.}
	\label{fig:preparation}
\end{figure}

We propose an excitation scheme for the dark states based on a Raman scheme through an intermediate state $\ket{s_{m_s}}$ with total angular momentum $F_s$, $m_s\in[-F_s,F_s]$, and decay rate $\Gamma_s$, see Fig.~\ref{fig:preparation}(a). We couple both states $g$ and $e$ off-resonantly to state $s$ with detuning $\Delta$ such that the effective Raman Hamiltonian is given by $\hat{H}_L^{\text{eff}} = \sum_{m,n} \left[ \Omega^{\text{eff}}_{mn}\, \hat\sigma^+_{mn} + \text{h.c.} \right]$ with $\hat\sigma^+_{mn}\equiv (\hat\sigma^-_{nm})^\dagger$ and the effective Rabi coupling $\Omega^{\text{eff}}_{mn} \equiv \Omega^\text{eff} \sum_{k} \tilde{C}_{m,k-m}^{(se)*}\, \tilde{C}_{n,k-n}^{(sg)}$.
Here, we defined $\tilde{C}^{(ab)}_{mq}\equiv \langle F_a,m;1,q | F_b,m+q \rangle (\mbf e^*_{q}\cdot\boldsymbol\epsilon_L^{(ab)})$ and $\Omega^\text{eff}\equiv\Omega^{(sg)}\Omega^{(se)}/\Delta$, where $\Omega^{(ab)}$ is the Rabi frequency of the laser connecting states $a$ and $b$ with polarization $\epsilon_L^{(ab)}$ ($a,b\in\{e,g,s\}$). This  is valid in the limit $|\Delta|\gg\Omega^{(sg)},\Omega^{(se)},\Gamma_s$.

The couplings $\Omega^{\text{eff}}_{mn}$ are, in general, different from $\Omega_{q}$, and they can lead to $\bra{D}\hat{H}^\text{eff}_L\ket{G}\neq0$.
To see this consider $F_g=F_e=F_s=1/2$ with both Raman lasers $\mbf e_z$-polarized.
Starting with both atoms in the ground state $\ket{G_{1/2}}\equiv\ket{g_{-1/2}g_{1/2}}$, the single photon Hamiltonian, $\hat{H}_L \propto \sum_{m} C^{q=0}_m \left[ \hat\sigma^+_{mm} + \text{h.c.} \right]$, couples to the superradiant state $\ket{S}\equiv \left( \ket{g_{1/2}\,e_{-1/2}} + \ket{g_{-1/2}\,e_{1/2}} \right)/\sqrt{2}$.
This results from the signs of $C^{q=0}_{1/2}=-C^{q=0}_{-1/2}$.
The Raman scheme essentially changes the signs of the couplings, $\hat{H}_L^{\text{eff}} \propto \sum_{m} (C^{q=0}_m)^2 \left[ \hat\sigma^+_{mm} + \text{h.c.} \right]$, which leads to a coupling to the dark state, $\hat{H}^\text{eff}_L\ket{G_{1/2}}\propto\ket{D_0}_{\{\frac{1}{2},\frac{1}{2}\}}$.

In general, the Raman Hamiltonian may couple the ground or dark states to undesired nondark states.
In the previous example, $\hat{H}_L^\text{eff}$ will couple the dark state to $\ket{ee}\equiv\ket{e_{-1/2}e_{1/2}}$.
To avoid populating any of these nondark states with decay rates $\propto\Gamma$ one can take advantage of the quantum Zeno effect. In the limit of small effective Rabi coupling, $\Omega^\text{eff}\ll\Gamma$, the fast decay acts as a continuous projection, and the nondark states are populated only at an effective rate $\sim(\Omega^\text{eff})^2/\Gamma$.
The nondark-state population can, thus, be suppressed by reducing $\Omega^\text{eff}/\Gamma$.
Figure~\ref{fig:preparation}(b) shows this for $F_g=F_e=1/2$, relevant for $^{171}$Yb. At large Rabi coupling, $\ket{ee}$ would get excited and decay into $\ket{S}$. At small Rabi coupling, however, high-contrast coherent oscillations between the ground and dark states can be observed.

Our Raman scheme can be used to excite dark states for other internal level structures too.
For this, one needs to choose (i) an initial ground-state $\ket{g_mg_n}$ and (ii) an intermediate state and laser polarizations such that the coupling to the dark state is nonzero.
Such ground states can in principle be prepared by optical pumping.
For example, to excite the dark state $\ket{D_0}_{\{F,F\}}$ of Eq.~(\ref{eq:dark_n2square}) for general $F_g=F_e=F$, one can start with $\ket{G_{F}}\equiv\ket{g_{-F}g_{F}}$ and choose $F_s= F$ with $\mbf e_z$-polarized lasers.
Figure~\ref{fig:preparation}(c) shows the coherent oscillations of the dark state obtained with this scheme for $F=9/2$, relevant for $^{87}$Sr. In this case, however, a full inversion into the dark state can not be obtained, since other ground states with $\ket{G_{F<9/2}}$ get coherently excited as well.
Further examples are given in the Supplementary Material~\cite{Supplemental}.

%%%%%%%%%%%%%					%%%%%%%%%%%%%
%%%%%%%%%%%%%		SECTION		%%%%%%%%%%%%%
%%%%%%%%%%%%%					%%%%%%%%%%%%%

{\it Implementation.---}
Alkaline-earth atoms are particularly well-suited due to their lack of hyperfine splitting in the ground state.
We consider, e.g., $^{171}$Yb or $^{87}$Sr with nuclear spins $I=1/2$ and $I=9/2$, respectively.

The natural choice would be to consider the clock transition $g= {}^1\text{S}_0$ to $e= {}^3\text{P}_0$, for which $F_g=F_e=I$ and $\Gamma\sim \text{mHz}$.
Because of its slow decay rate, however, dark states on this transition will be currently hard to observe, due to other competing decay mechanisms such as light scattering~\cite{HutsonYe2019}.
A better alternative would be to use the faster decaying transition to $e= {}^3\text{P}_1$, $\Gamma\sim\text{kHz}$, either with $F_e=I$ or with $F_e=I+1$.
To implement the dark state preparation scheme one could use $s= {}^1\text{P}_1$~\cite{SantraGreenePRL94}, $s= {}^1\text{D}_2$, or $s= {}^3\text{D}_2$ as an intermediate state.

We anticipate two possible sources of complications: collisions and stray magnetic fields.
In general, collisions may induce mixing of dark and nondark states, and thus reduce the lifetime of the dark state.
However, by tightly trapping the atoms so that they are confined to be in the motional ground state, the number of possible collision channels is greatly reduced.
In particular, conservation of total angular momentum implies that the $\ket{F_T,M}$ dark eigenstates (\ref{eq:dark_n2square}) and (\ref{eq:dark_n2V}) should be good collisional eigenstates, as long as collisions are dominated by the long-range part of the potential.
Depolarization effects in $^3\text{P}_1$ due to magnetic dipole interactions can be avoided if one works in the lowest energy $F$ state.

Nonvanishing stray magnetic fields lead to Zeeman splittings $\Delta_z$ that also cause mixing between dark and nondark states.
Assuming magnetic field fluctuations in optical lattice experiments to be of order $\mu\text{G}$~\cite{OelkerYe2019} implies that for the $^3$P$_1$ linear Zeeman shift of order $\text{MHz}/\text{G}$, one has $\Delta_z/\Gamma\sim10^{-3}\ll1$.
In this case, the quantum Zeno effect again protects the dark state, which acquires an effective decay rate given by $\Gamma_\text{eff}\sim\Delta_z^2/\Gamma$.
Figure~\ref{fig:preparation}(d) shows an example, for $F_g=F_e=1/2$, of a full Ramsey sequence where the $\ket{e_{\pm1/2}}$ states are detuned by $\pm\Delta_z/2$. The decay implies a reduced contrast for phase estimation.
The lifetime of the dark state is further limited by the decay to the next motional band with $\Gamma_\text{eff}\sim\eta^2\Gamma$, for small Lamb-Dicke parameter $\eta$~\cite{ZollerSandnerPRA2011}.

%%%%%%%%%%%%%					%%%%%%%%%%%%%
%%%%%%%%%%%%%		SECTION		%%%%%%%%%%%%%
%%%%%%%%%%%%%					%%%%%%%%%%%%%

{\it Applications.---}We have shown that multilevel fermions loaded into doubly-filled optical lattices support a large set of dark states and, thus, offer an attractive platform to experimentally observe strong subradiance.
Superpositions of ground and dark states can be useful for building precise optical clocks, even on internal levels that are not naturally long lived.
The multilevel dark states will not suffer from dipole interaction shifts, which may constitute a fundamental limit for  3D lattice clocks. Qubits made of ground and dark states will  form a  large decoherence-free subspace which could be used in quantum information science, e.g.~as quantum memories, as well as for quantum simulation, e.g.~by including superexchange interactions, or for quantum optical devices, e.g.~to create mirrors or interesting photonic states.

\begin{acknowledgments}
We thank D.~Barberena and I.~Kimchi for their insights into the role of total angular momenta eigenstates, as well as P.~Julienne, J.~P.~D'Incao, C.~Sanner, J.~Ye, A.~Kaufman, and M.~Perlin for useful discussions on the topic.
This work is supported by the AFOSR grant FA9550-18-1-0319 and its MURI Initiative, by the DARPA and ARO grant W911NF-16-1-0576,  the ARO single investigator award W911NF-19-1-0210,  the NSF PHY1820885, NSF JILA-PFC PHY-1734006 grants, and by NIST.
\end{acknowledgments}

\bibliography{darks_multilevel_pauli_bibliography}

%%%%%%%%%%%%%							%%%%%%%%%%%%%
%%%%%%%%%%%%%		SUPPLEMENTARY		%%%%%%%%%%%%%
%%%%%%%%%%%%%							%%%%%%%%%%%%%

\vfill
\newpage

\onecolumngrid
\vspace{\columnsep}
\begin{center}
\textbf{\large Supplementary Material: Dark states of multilevel fermionic atoms in doubly-filled optical lattices}
\end{center}
\vspace{\columnsep}
\twocolumngrid

\setcounter{equation}{0}
\setcounter{figure}{0}
\setcounter{table}{0}
\setcounter{page}{1}
\makeatletter
\renewcommand{\theequation}{S\arabic{equation}}
\renewcommand{\thefigure}{S\arabic{figure}}
\renewcommand{\bibnumfmt}[1]{[S#1]}
\renewcommand{\citenumfont}[1]{#1}

%%%%%%%%%%%%%					%%%%%%%%%%%%%
%%%%%%%%%%%%%		SECTION		%%%%%%%%%%%%%
%%%%%%%%%%%%%					%%%%%%%%%%%%%

\section{Dipole matrix elements}

We separate the matrix elements of the dipole operator $\hat{\mbf d}$ using the Wigner-Eckart theorem~\cite{brown2003rotational} as
\begin{equation}
	\langle e_m | \hat{\mbf d} | g_n \rangle \equiv \frac{ \mbf d^{(eg)}_{mn}\,d^\text{rad}_{ge}}{\sqrt{2F_e+1}},
\label{eq:d_def}
\end{equation}
where $\mbf d^{(eg)}_{mn}$ is the spherical and $d^\text{rad}_{ge}$ the radial part.
The relationship between the spherical part and the Clebsch-Gordan coefficients is given by ($q=0,\pm1$)
\begin{equation}
	\mbf d^{(eg)}_{m+q,m} = \langle F_g,m;1,q |F_e,m+q \rangle\, \mbf e_{q}^*.
\end{equation}

%%%%%%%%%%%%%					%%%%%%%%%%%%%
%%%%%%%%%%%%%		SECTION		%%%%%%%%%%%%%
%%%%%%%%%%%%%					%%%%%%%%%%%%%

\section{Onsite dipole interaction coefficients}

To compute the dipole interaction between two atoms occupying the same site, the spatial part of the wave-function needs to be taken into account as
\begin{align}
	G^{ii} = \int d\mbf r\, d\mbf r'\, G(\mbf r-\mbf r')\, |w_i(\mbf r)|^2\, |w_i(\mbf r')|^2.
\end{align}
Here, $G(\mbf r-\mbf r')$ is the dyadic Green's tensor defined in the main text, and $w_i(\mbf r)=w(\mbf r-\mbf r_i)$ is the Wannier function of the lowest motional eigenstate of lattice site $i$. In the limit of a deep trap, $w_i(\mbf r)$ can be approximated by the ground-state wave-function of a harmonic potential, which is a Gaussian profile. The leading-order terms of $G(\mbf r)$ for short distances are given by
\begin{align}
	\Re G(\mbf r) \approx&\,  -\frac{3\Gamma}{4} \frac{ 1-3\hat{r} \otimes \hat{r} }{ (k_0r)^3 } ,\\
	\Im G(\mbf r) \approx&\, \frac{\Gamma}{2} \mathbb{1}.
\end{align}
This implies $\Im G^{ii}=\frac{\Gamma}{2}\mathbb{1}$.
For a radially symmetric potential, the spherical integral over $\Re G(\mbf r)$ vanishes, and hence $\Re G^{ii}=0$.

\begin{figure}[b!]
\centering
	\includegraphics[width=\columnwidth]{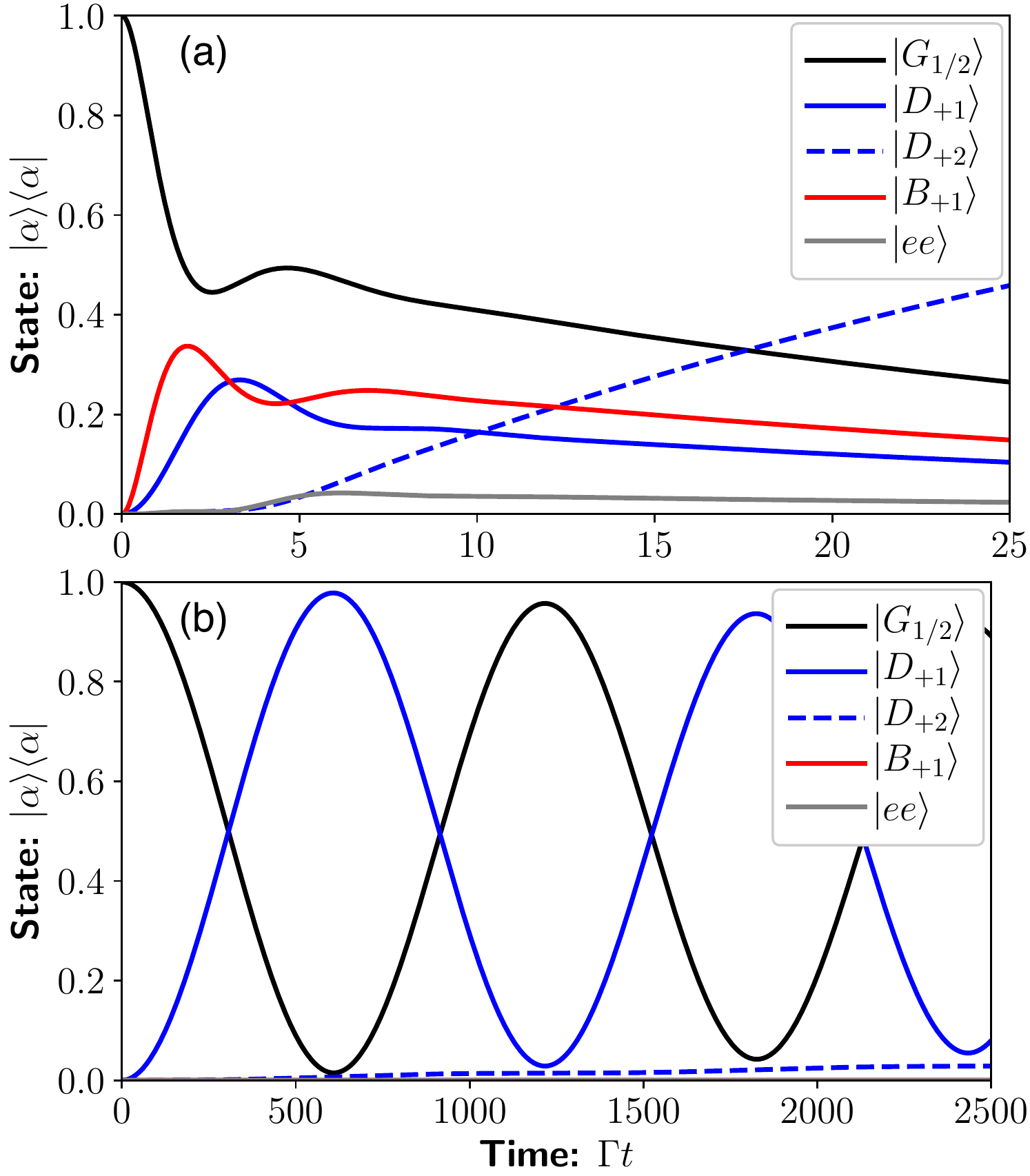}
	\caption{Raman preparation scheme applied to $F_g=1/2$, $F_e=3/2$ for (a) $\Omega^\text{eff}=\Gamma$ and (b) $\Omega^\text{eff}=0.01\Gamma$.}
	\label{fig:SMramanYb2-4}
\end{figure}

%%%%%%%%%%%%%					%%%%%%%%%%%%%
%%%%%%%%%%%%%		SECTION		%%%%%%%%%%%%%
%%%%%%%%%%%%%					%%%%%%%%%%%%%

\section{Preparation}

We provide here an example of the Raman excitation scheme applied to the internal level structure $F_e=F_g+1$. Specifically, we consider $F_g=1/2\rightarrow F_e=3/2$ and show how to prepare the dark state $\ket{D_{+1}}_{\{1/2,3/2\}}$ starting from the ground state $\ket{G_{1/2}}\equiv\ket{g_{-1/2}g_{1/2}}$. For this we use an intermediate state with $F_s=3/2$ and choose the lasers to have polarizations $\epsilon^{(sg)}_L=\mbf e_+$ and $\epsilon^{(se)}_L=\mbf e_z$.

Figure~\ref{fig:SMramanYb2-4} shows the evolution of the occupations of different states for different effective Rabi couplings $\Omega^\text{eff}$. For large effective Rabi coupling $\Omega^\text{eff}=\Gamma$ [Fig.~\ref{fig:SMramanYb2-4}(a)] the laser substantially excites both the target dark state $\ket{D_{+1}}_{\{1/2,3/2\}}=\frac{\sqrt{3}}{2}\ket{g_{1/2}e_{1/2}}+\frac{1}{2}\ket{g_{-1/2}e_{3/2}}$ (blue line), as well as the bright state orthogonal to it, $\ket{B_{+1}}_{\{1/2,3/2\}}\equiv\frac{1}{2}\ket{g_{1/2}e_{1/2}}-\frac{\sqrt{3}}{2}\ket{g_{-1/2}e_{3/2}}$ (red line).
These states subsequently couple to the doubly-excited state $\ket{ee}\equiv\ket{e_{1/2}e_{3/2}}$ (gray line), which gets excited too.
Part of $\ket{ee}$ then decays into the maximally stretched dark state $\ket{D_{+2}}_{\{1/2,3/2\}}=\ket{g_{1/2}e_{3/2}}$ (dashed blue line).

The occupancy of the bright and doubly-excited states can be significantly reduced using the quantum Zeno effect, as explained in the main text. Figure~\ref{fig:SMramanYb2-4}(b) shows for $\Omega^\text{eff}=0.01\Gamma$ high-contrast coherent oscillations between the ground and target dark state, with a negligible occupation of the nondark states.

At long times, the atoms slowly end up in the maximally stretched dark state $\ket{D_{+2}}$, which neither decays nor couples to other states for the lasers chosen. Thus, this can be seen as a way of preparing such maximally stretched states in an optical pumping fashion.

\end{document}